\documentclass{article} 
\usepackage{amsmath,amssymb,amsfonts}   
\usepackage{graphicx,bm}

 \usepackage{tikz-cd}

\usepackage{bm,braket}

\newcommand{\calU}{{\mathcal U}}

\newcommand{\calP}{{\mathcal P}}

\newcommand{\calF}{{\mathcal F}}

\newcommand{\R}{{\mathbb R}}
\renewcommand{\L}{{\mathbb L}}
\newcommand{\X}{\mathbf{X}}
\renewcommand{\P}{\mathbb{P}}
\newcommand{\PP}{\widetilde{P}}

\newcommand{\x}{\mathbf{x}}

\newcommand{\e}{{\mathrm e}}
\newcommand{\E}{{\mathbb E}}

\newcommand{\n}{\mathbf n}

\renewcommand{\P}{\mathbb P}
\newcommand{\p}{\widetilde{p}}

\newcommand{\U}{\widetilde{U}}

\newcommand{\ellh}{\widehat{\ell}}

\begin{document}
\title{Accumulation times for diffusion-mediated surface reactions}
\author{ \em
P. C. Bressloff, \\ Department of Mathematics, 
University of Utah \\155 South 1400 East, Salt Lake City, UT 84112}

\maketitle

\begin{abstract} In this paper we consider a multiparticle version of a recent probabilistic framework for studying diffusion-mediated surface reactions. The basic idea of the probabilistic approach is to consider the joint probability density or generalized propagator for particle position and the so-called boundary local time. The latter characterizes the amount of time that a Brownian particle spends in the neighborhood of a totally reflecting boundary; the effects of surface reactions are then incorporated via an appropriate stopping condition for the local time. The propagator is determined by solving a Robin boundary value problem, in which the constant rate of reactivity is identified as the Laplace variable $z$ conjugate to the local time, and then inverting the solution with respect to $z$. Here we reinterpret the propagator as a particle concentration in which surface absorption is counterbalanced by particle source terms. We investigate conditions under which there exists a non-trivial steady state solution, and analyze the relaxation to steady state by calculating the corresponding accumulation time. In particular, we show that the first two moments of the stopping local time density have to be finite.

 \end{abstract}

\section{Introduction}

The classical diffusion equation $\partial u(\x,t) /\partial t=D\nabla^2 u(\x,t) $ for particle concentration $u$ in a bounded domain $\Omega$ can be interpreted as a conservation law describing the flux of many particles moving from regions of high concentration to regions of low concentration at a rate that depends on the local concentration gradient (Fick's law). In the case of a partially absorbing boundary $\partial \Omega$ with constant reactivity $\kappa_0$ and unit outward normal $\n$, the corresponding boundary condition is of the Robin form $D\nabla u(\x,t)\cdot \n+\kappa_0 u(\x,t)=0$ for all $\x\in \partial \Omega$. Dirichlet and Neumann boundary conditions are obtained in the limits $\kappa_0\rightarrow \infty$ and $\kappa_0\rightarrow 0$, respectively. If $\kappa_0>0$ then the concentration vanishes in the large-time limit. On the other hand, if the boundary $\partial \Omega$ is partitioned according to $\Omega=\partial \Omega_1\cup \Omega_2$ with a Robin condition on $\partial \Omega_2$ and a constant flux condition $D\nabla u\cdot \n =J_0$ on $\Omega_1$, then there exists a non-trivial steady-state concentration $u^*(\x)=\lim_{t\rightarrow\infty}u(\x,t)$. Moreover, the relaxation to steady-state can be determined by treating the fractional deviation from the steady-state concentration as a cumulative distribution whose mean is identified with the so-called local accumulation time. In contrast to a global measure of the relaxation rate based on the principal nonzero eigenvalue of the Laplacian, the accumulation time takes into account the fact that different spatial regions can relax at different rates. Accumulation times were originally used to estimate the time to form a protein concentration gradient during morphogenesis \cite{Berez10,Berez11,Gordon11}, but have subsequently been applied to a wider range of diffusion processes, including intracellular protein gradient formation \cite{Bressloff19}, search processes with stochastic resetting \cite{Bressloff21C}, and gap junctions \cite{Bressloff22}.

It is well known that one can also formulate diffusion at the single-particle level by considering the probability density $p(\x,t|\x_0)$ for the random position $\X_t $ of the particle at time $t$ given the initial position $\X_0=\x_0$. The evolution equation for the probability density in a bounded domain $\Omega$ is identical to the macroscopic diffusion equation with homogeneous boundary conditions and the additional constraint $\int_{\Omega} p(\x,t)d\x \leq 1$ for all $t>0$. Indeed, one way to recover the macroscopic version is to consider a large population of $N$ independently diffusing particles and to set $u(\x,t)=\int_{\Omega}u(\x_0) p(\x,t|\x_0)d\x_0$ where $u(\x_0)$ is the initial concentration. Individual trajectories of a single particle are generated by a stochastic differential equation (SDE) that, in the case of pure diffusion in $\R^d$ is given by a Wiener process. However, incorporating boundary conditions into the underlying SDE is non-trivial. In the case of a totally absorbing boundary (Dirichlet), one simply stops the Brownian motion on the first encounter between particle and boundary; the random time at which this event occurs is known as the first passage time (FPT). On the other hand, in the case of a totally reflecting boundary it is necessary to modify the stochastic process itself by introducing a Brownian functional known as the boundary local time \cite{Levy39,McKean75,Freidlin85,Majumdar05}. The latter determines the amount of time that a Brownian particle spends in the neighborhood of points on the boundary. Probabilistic versions of the Robin boundary condition can also be constructed \cite{Papanicolaou90,Milshtein95,Singer08}.

One of the interesting features of the single-particle perspective is that one can consider microscopic models of absorption that go beyond the constant reactivity models underlying the Robin boundary condition. One such probabilistic framework has recently been developed by Grebenkov \cite{Grebenkov19a,Grebenkov19b,Grebenkov20,Grebenkov22}, who considers the joint probability density or generalized propagator $P(\x,\ell,t)$ for the pair $(\X_t,\ell_t)$ in the case of a perfectly reflecting boundary $\partial \Omega$, where $\X_t$ and $\ell_t$ denote the particle position and local time, respectively. Partial absorption is then introduced by terminating the diffusion process at the stopping time 
$
{\mathcal T}=\inf\{t>0:\ \ell_t >\widehat{\ell}\}$,
where $\widehat{\ell}$ is a randomly distributed local time threshold with probability distribution $\Psi(\ell) = \P[\ellh>\ell]$. The marginal probability density for particle position is then defined according to
 $  p(\x,t)=\int_0^{\infty} \Psi(\ell)P(\x,\ell,t)d\ell$. A crucial observation is that the classical Robin boundary condition for the diffusion equation corresponds to the exponential distribution
$\Psi(\ell) =\e^{-\gamma \ell}$, where $\gamma =\kappa_0/D$. This implies that one can obtain the generalized propagator $P(\x,\ell,t)$ by Laplace transforming with respect to $\ell$, solving the resulting Robin boundary value problem (BVP) with the corresponding Laplace variable $z$ acting as a constant reactivity, and then calculating the inverse Laplace transform. 

The general theory of single-particle diffusion in domains with partially absorbing surfaces motivates developing the analogous theory at the macroscopic level of multiparticle diffusion, which is the subject of the current paper. We proceed by reinterpreting the generalized propagator as a generalized concentration $\calU(\x,\ell,t)$ with an associated marginal concentration $u(\x,t)=\int_0^{\infty}\Psi(\ell) \calU(\x,\ell,t)d\ell$. This allows us to include additional source terms that counteract the loss of particles due to absorption. We then investigate conditions under which there exists a non-trivial steady state solution, and analyze the relaxation to steady state by calculating the corresponding accumulation time. The structure of the paper is as follows. In section 2 we briefly review the single-particle theory for a general bounded domain $\Omega \subset \R^d$. We formulate the multiparticle version in section 3, and solve the resulting Robin BVP in Laplace space using the spectral decomposition of a Dirichlet-to-Neumann operator. This allows us to invert the Laplace transform with respect to the local time and identify necessary conditions for the existence of a steady-state solution and the associated accumulation time. These conditions require that various moments of the stopping local time density are finite. Finally, we illustrate the theory by considering diffusion in a finite interval (section 4) and in a spherical shell (section 5), where the Robin BVP can be solved explicitly without recourse to spectral theory.

  \section{Partially absorbing surfaces and the local time propagator}
  
  In this section we describe the encounter-based method for analyzing single-particle diffusion in domains with partially absorbing surfaces. Our presentation is equivalent to previous versions \cite{Grebenkov20,Bressloff22a}, but is developed in a form that is easily generalizable to the multiparticle case. In particular, we focus on the BVP for the local time propagator and its double Laplace transform.
  
 Consider a particle diffusing in the bounded domain $\Omega$ with a partially absorbing boundary $\partial \Omega$. Let $\X_t \in \Omega $ represent the position of the particle at time $t$. For the moment suppose that $\partial \Omega$ is totally reflecting. The boundary local time is defined according to \cite{Levy39,McKean75,Majumdar05,Grebenkov19a}
\begin{equation}
\label{loc}
\ell_t=\lim_{\delta\rightarrow 0} \frac{D}{\delta} \int_0^tH(\delta-\mbox{dist}(\X_{\tau},\partial \Omega))d\tau,
\end{equation}
where $H$ is the Heaviside function. Note that although $\ell_t$ has units of length due to the additional factor of $D$, it essentially specifies the amount of time that the particle spends in an infinitesimal neighborhood of the surface $\partial \Omega$. 
Let $P(\x,\ell,t|\x_0)$ denote the joint probability density or propagator for the pair $(\X_t,\ell_t)$. The propagator satisfies a BVP that can be derived using integral representations \cite{Grebenkov20} or path-integrals \cite{Bressloff22a}:
\begin{subequations}
\label{Ploc}
\begin{align}
  &\frac{\partial P(\x,\ell,t|\x_0)}{\partial t}=D\nabla^2 P(\x,\ell,t|\x_0) , \quad P(\x,\ell,0|\x_0)=\delta(\x-\x_0)\delta(\ell),\ \x \in \Omega,\\
 &-D\nabla P(\x,\ell,t|\x_0) \cdot \n= D P(\x,\ell=0,t|\x_0) \ \delta(\ell)  +D\frac{\partial}{\partial \ell} P(\x,\ell,t|\x_0) \mbox{ for }  \x\in \partial \Omega, \\
 &P(\x,\ell=0,t|\x_0)=-\nabla p_{\infty}(\x,t|\x_0)\cdot \n \mbox{ for } \x\in \partial \Omega, 
\end{align}
\end{subequations}
where $p_{\infty}$ is the probability density in the case of a totally absorbing boundary: 
\begin{subequations}
\label{pinf}
\begin{align}
	&\frac{\partial p_{\infty}(\x,t|\x_0)}{\partial t} = D\nabla^2 p_{\infty}(\x,t|\x_0),
&
 \\
 & \quad  p_{\infty}(\x,0|\x_0)=\delta(\x-\x_0),\ \x\in \Omega,\quad p_{\infty}(\x,t|\x_0)=0,\  \x\in \partial \Omega.
	\end{align}
	\end{subequations}
	An intuitive interpretation of the boundary condition (\ref{Ploc}b) is that the rate at which the local time increases is proportional to the flux into the boundary when $\ell_t >0$. However, this process only starts once the particle has reached the surface for the first time, which is identical to the case of a totally absorbing surface.

The BVP (\ref{Ploc}) can be solved by introducing the double Laplace transform 
\begin{equation}
 \label{dLT}
 \calP(\x,z,s)\equiv \int_0^{\infty}\e^{-z\ell}\int_0^{\infty}\e^{-st}P(\x,\ell,t)dtd\ell,
 \end{equation} 
with \cite{Grebenkov20,Grebenkov22,Bressloff22a}
\begin{subequations}
\label{PlocLT}
\begin{align}
 &D\nabla^2 \calP(\x,z,s|\x_0)-s\calP(\x,z,s|\x_0)=-\delta(\x-\x_0),\ \x \in \Omega,\\
&-\nabla \calP(\x,z,s|\x_0) \cdot \n=z\calP(\x,z,s|\x_0) \mbox{ for }  \x\in \partial \Omega .
\end{align}
\end{subequations}
Suppose that $z\equiv \gamma_0=\kappa_0/D$ for some constant $\kappa_0$ and define the marginal density $\p(\x,s|\x_0)=\calP(\x,\kappa_0/D,s|\x_0)$. 
The BVP (\ref{PlocLT}) for $\p(\x,s|\x_0)$ is then identical to the $s$-Laplace transformed diffusion equation in the case of a Robin boundary condition on $\partial \Omega$ with a constant rate of reactivity $\kappa_0$. In order to further understand this result, we follow along the lines of Ref. \cite{Grebenkov19b,Grebenkov20,Grebenkov22} by introducing the absorption stopping time 
${\mathcal T}=\inf\{t>0:\ \ell_t >\widehat{\ell}\}$
 with $\widehat{\ell}$ an exponentially distributed random variable that represents a stopping local time. That is, $\P[\widehat{\ell}>\ell]=\e^{-\gamma_0\ell}$. Given that $\ell_t$ is a nondecreasing process, the condition $t < {\mathcal T}$ is equivalent to the condition $\ell_t <\widehat{\ell}$. Define the marginal density $p(\x,t|\x_0)$ as
\begin{align*}
p(\x,t|\x_0)d\x&\equiv\P[\X_t \in (\x,\x+d\x), \ \ell_t < \widehat{\ell}|\X_0=\x_0]\\
&=\int_0^{\infty} d\ell \ \gamma_0\e^{-\gamma_0\ell}\P[\X_t \in (\x,\x+d\x), \ \ell_t < \ell |\X_0=\x_0]\\
&=\int_0^{\infty} d\ell \ \gamma_0 \e^{-\gamma_0\ell}\int_0^{\ell} d\ell' [P(\x,\ell',t|\x_0)d\x].
\end{align*}
Using the identity
\[\int_0^{\infty}d\ell \ f(\ell)\int_0^{\ell} d\ell' \ g(\ell')=\int_0^{\infty}d\ell' \ g(\ell')\int_{\ell'}^{\infty} d\ell \ f(\ell)\]
for arbitrary integrable functions $f,g$, we have
\begin{equation}
\label{bob}
p(\x,t|\x_0)=\int_0^{\infty} \e^{-\gamma_0\ell}P(\x,\ell,t|\x_0)d\ell.
\end{equation}
Laplace transforming with respect to $t$ immediately establishes that $\p(\x,s|\x_0)$ satisfies the BVP (\ref{PlocLT}) with $z=\gamma_0$. Hence, the Robin boundary condition is equivalent to an exponential law for the stopping local time $\widehat{\ell}_t$.

The advantage of formulating the Robin boundary condition in terms of the generalized propagator is that one can consider a more general probability distribution $\Psi(\ell) = \P[\ellh>\ell]$ for the stopping local time $\ellh$ such that \cite{Grebenkov19b,Grebenkov20,Grebenkov22}
  \begin{equation}
  \label{oo}
  p (\x,t|\x_0)=\int_0^{\infty} \Psi(\ell)P(\x,\ell,t|\x_0)d\ell \ \mbox{ for } \x \in \Omega .
  \end{equation}
This accommodates a much wider class of surface reactions where, for example, the reactivity $\kappa(\ell)$ depends on the local time $\ell$ (or the number of surface encounters):
\begin{equation}
\label{kaell}
\Psi(\ell)=\exp\left (-\frac{1}{D}\int_0^{\ell}\kappa(\ell')d\ell'\right ).
\end{equation}
 Laplace transforming equation (\ref{oo}) with respect to $t$ gives
  \begin{equation}
  \label{oo2}
  \p  (\x,s|\x_0)=\int_0^{\infty} \Psi(\ell){\mathcal L}_{\ell}^{-1}[\calP(\x,z,s|\x_0)] d\ell \ \mbox{ for } \x \in \Omega ,
  \end{equation}
  where $\calP(\x,z,s|\x_0)$ is the solution of the Robin BVP given by equations (\ref{PlocLT}). That is, the marginal density $\p(\x,s|\x_0)$ for a general distribution $\Psi(\ell)$ can be obtained by solving a classical Robin BVP with effective reactivity $\kappa=zD$ and then inverting the Laplace transform with respect to $z$.  
   
One important quantity of interest that can be obtained directly from $ \p(\x,s|\x_0)$ is the MFPT for absorption. In order to show this, consider the probability flux
\begin{equation}
J(\x_0,t)=-D\int_{\partial \Omega} \nabla p (\x,t|\x_0) \cdot \n \, d\x.
\end{equation}
Multiplying both sides of equation (\ref{Ploc}b) with respect to the stopping local time distribution $\Psi(\ell)$ and integrating by parts with respect to $\ell$ gives
\begin{equation}
 \int_0^{\infty}\Psi(\ell) \nabla P(\x,\ell,t|\x_0) \cdot \n \, d\ell =-\int_0^{\infty}\psi(\ell) P(\x,\ell,t|\x_0)d\ell,
\end{equation}
where $\psi(\ell)=-\Psi'(\ell)$. Integrating both sides with respect to $\x\in \partial \Omega$ and using equation (\ref{oo}) then implies that
\begin{equation}
J(\x_0,t)=D\int_{\partial \Omega}\left [\int_0^{\infty}\psi(\ell)P(\x,\ell,t|\x_0)d\ell \right ]d\x.
\end{equation}
Finally, Laplace transforming with respect to time $t$ yields the result
\begin{equation}
\label{JLT}
\widetilde{J}(\x_0,s)=D\int_{\partial \Omega}\left [\int_0^{\infty}\psi(\ell){\mathcal L}_{\ell}^{-1}[\calP(\x,z,s|\x_0)]d\ell \right ]d\x.
\end{equation}
Finally, note that the flux $J(\x_0,t)$ can be identified with the first passage time (FPT) density for absorption. In particular, the MFPT is
\begin{align}
\label{MFPT}
T(\x_0)\equiv \E[{\mathcal T}] =\int_0^{\infty} tJ(\x_0,t)dt =-\left . \frac{\partial \widetilde{J}(\x_0,s)}{\partial s}\right |_{s=0}.
\end{align}
Similarly, higher-order moments of the FPT density can be expressed in terms of higher-order derivatives of the Laplace transformed flux $ \widetilde{J}(\x_0,s)$.

  \section{Multiparticle interpretation and the accumulation time}
  
\begin{figure}[t!]
\centering
  \includegraphics[width=6cm]{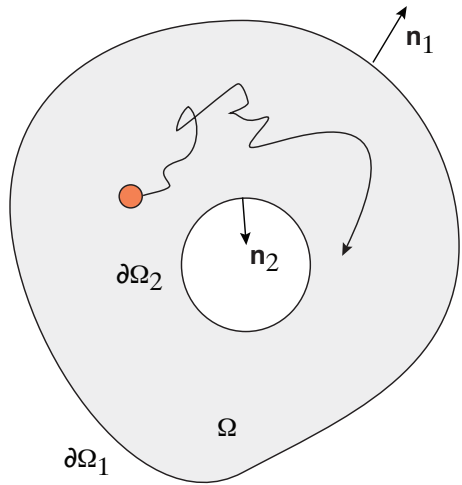}
  \caption{Diffusion of a particle in a bounded domain $\Omega$ with $\partial \Omega =\partial \Omega_1\cup \partial \Omega_2$. There is a constant flux $J_0$  through the boundary $\partial \Omega_1$, whereas $\partial \Omega_2$ is partially absorbing. The probability of particle absorption depends on the amount of time each particle spends in a neighborhood of $\partial \Omega_2$, which is specified by the local accumulation time $\ell_t$. Note that the unit normals $\n_{1,2}$ are directed towards the exterior of $\Omega$.}
  \label{fig1}
\end{figure}

 We now consider a multiparticle version of diffusion-mediated surface absorption in which the propagator $P(\x,\ell,t|\x_0)$ is reinterpreted as the concentration $U(\x,\ell,t|\x_0)$ of a large population of independently diffusing particles that are initially localized at $\X=\x_0$ and $\ell=0$. One difference from the single particle BVP is that we can now include source terms, either within the bulk domain or in part of the boundary, resulting in a non-trivial steady-state 
  \begin{equation}
  U^*(\x,\ell)=\lim_{t\rightarrow \infty} U(\x,\ell,t|\x_0)=\lim_{s\rightarrow 0}s \U(\x,\ell,s|\x_0).
  \end{equation}
  As a concrete example suppose that the boundary of the domain $\Omega$ is partitioned into two separate boundaries, $\partial\Omega=\partial \Omega_1\cup \partial \Omega_2$, where $\partial \Omega_2$ is partially absorbing whereas a constant flux condition is imposed on $\partial \Omega_1$, see Fig. \ref{fig1}. The unit normals to the surfaces $\partial\Omega_{1,2}$ are denoted by $\n_{1,2}$, and are directed outwards from the interior of $\Omega$. The multiparticle version of the BVP (\ref{Ploc}) is taken to be
  \begin{subequations}
\label{mPloc}
\begin{align}
  &\frac{\partial U(\x,\ell,t)}{\partial t}=D\nabla^2 U(\x,\ell,t), \quad U(\x,\ell,0)=0, \ \x \in \Omega ,\\
 &-D\nabla U(\x,\ell,t) \cdot \n_2= D U(\x,\ell=0,t) \ \delta(\ell)  +D\frac{\partial}{\partial \ell} U(\x,\ell,t) ,\  \x\in \partial \Omega_2, \\
 & D\nabla U(\x,\ell,t) \cdot \n_1 =J_0\delta(\ell),\ \x \in \partial \Omega_1,\\
 &U(\x,\ell=0,t)=-\nabla u_{\infty}(\x,t)\cdot \n_2 \mbox{ for } \x\in \partial \Omega, 
\end{align}
\end{subequations}
where $u_{\infty}$ is the concentration in the case of a totally absorbing boundary $\partial \Omega_1$. In contrast to the single-particle case, we assume that the domain $\Omega$ does not initially contain any particles.
Performing a double Laplace transform then yields the multiparticle version of the BVP (\ref{PlocLT}):
\begin{subequations}
\label{mPlocLT}
\begin{align}
 &D\nabla^2 \calU(\x,z,s)-s\calU(\x,z,s)=0,\ \x \in \Omega,\\
&-\nabla \calU(\x,z,s) \cdot \n_2=z\calU(\x,z,s) \mbox{ for }  \x\in \partial \Omega_2, \\
&D\nabla \calU(\x,z,s) \cdot \n_1=\frac{J_0}{s} \mbox{ for }  \x\in \partial \Omega_1.
\end{align}
\end{subequations}

\subsection{Dirichlet-to Neumann operator}

It is well known from classical PDE theory that the solution of a general Robin BVP can be computed in terms of the spectrum of a Dirichlet-to-Neumann operator. This was applied to the single-particle propagator BVP in Ref. \cite{Grebenkov20}. Here we consider the analogous result for the BVP (\ref{mPlocLT}). The basic idea is to replace the Robin boundary condition (\ref{mPlocLT}b) by the Dirichlet condition $\calU(\x,s)=f(\x,s),\  \x\in \partial \Omega_2$ and to find the function $f$ for which $\calU$ is also the solution to the original BVP. (For the moment we drop the explicit dependence on $z$.) 
The solution of the simplified BVP can be written in the form
\begin{equation}
\label{mcalF}
 \calU(\x,s)=-D\int_{\partial \Omega_2}f(\x',s)\nabla_{\x'}G(\x',s|\x)\cdot \n_2\, d\x'+\frac{J_0}{s} \int_{\partial \Omega_1} G(\x',s|\x)d\x'.
\end{equation}
where $G(\x,s|\x_0)$ denotes the Green's function for the modified Helmholtz equation,
\begin{align}
\label{nabG}
 &D\nabla^2 G(\x,s|\x_0)-sG(\x,s|\x_0)=-\delta(\x-\x_0),\ \x \in \Omega,
 \end{align}
 with boundary conditions $G(\x,s|\x_0)=0$ for all $\x\in \partial \Omega_2$ and $\nabla G(\x,s|\x_0)\cdot \n_1=0$ for all $\x\in \partial \Omega_1$. The unknown function $f$ satisfies the equation
\begin{align}
\label{mfH0}
[\L_s f](\x,s)+zf(\x,s)=g(\x,s), \quad \x \in \partial \Omega_2,
\end{align}
where $\L_s$ is the Dirichlet-to-Neumann operator on $\Omega_2$,
\begin{equation}
\label{mfH}
[\L_s f](\x,s)=-D\partial_{\sigma} \int_{\partial \Omega_2}f(\x',s)\partial_{\sigma'} G(\x',s|\x)d\x',
\end{equation}
and
\begin{equation}
g(\x,s)=-\frac{J_0}{s} \partial_{\sigma} \int_{\partial \Omega_1} G(\x',s|\x)d\x'.
\end{equation}
In the above equations we have set  $\partial_{\sigma}=\n_2\cdot \nabla_{\x}$ and  $\partial_{\sigma'}=\n_2'\cdot \nabla_{\x'}$.

When the surface $\partial \Omega_2$ is bounded, the Dirichlet-to-Neumann operator $\L_s$ has a discrete spectrum, that is, there exists a countable set of eigenvalues $\mu_n(s)$ and eigenfunctions $v_n(\x,s)$ satisfying (for fixed $s$)
\begin{equation}
\label{eig}
\L_s v_n(\x,s)=\mu_n(s)v_n(\x,s).
\end{equation}
It can be shown that the eigenvalues are non-negative and that the eigenfunctions form a complete orthonormal basis in $L_2(\partial \Omega_2)$. We can now solve equation (\ref{mfH0}) for $f$ by introducing the eigenfunction expansion
\begin{equation}
\label{eig2}
f(\x,s)=\sum_{m=0}^{\infty}f_m(s) v_m(\x,s).
\end{equation}
Substituting into (\ref{mfH}) and taking the inner product with the adjoint eigenfunction $v_n^*(\x,s)$ determines the coefficients $f_n$ in terms of the $n$th coefficient in the corresponding expansion of $g(\x,s)$:
\begin{equation}
\label{specf}
f_n(s)=\frac{g_n(s)}{\mu_n(s)+z},\quad g_n(s)=\int_{\partial \Omega_2}v_n^*(\x,s) g(\x,s)d\x.
\end{equation}
Substituting for $f(\x,s)$ in equation (\ref{mcalF}), we have
 \begin{equation}
 \label{mspec2}
  \calU(\x,z.s)=\frac{J_0}{sD}\sum_{n=0}^{\infty} \frac{1}{\mu_n(s)+z} \Delta_n^*(s) {\mathcal V}_n(\x,s),
  \end{equation}
  with
\begin{equation}
{\mathcal V}_n(\x,s)=-D\int_{\partial \Omega_2} v_n(\x',s)\partial_{\sigma'}G(\x',s|\x)d\x'
\end{equation}  
and
\begin{equation}
\Delta^*_n(s) =-D\int_{\partial \Omega_2} v^*_n(\x,s)\partial_{\sigma}\left [\int_{\partial \Omega_1}G(\x',s|\x)d\x'\right ]d\x.
\end{equation}  
Finally, inverting with respect to $z$ yields
  \begin{equation}
  \label{solU}
  \U(\x,\ell,s)= \frac{J_0}{sD}\sum_{n=0}^{\infty} \Delta_n^*(s){\mathcal V}_n(\x,s) \e^{-\mu_n(s) \ell}.
  \end{equation}
  In the following we will assume that the infinite series (\ref{solU}) is uniformly convergent so that we can reverse the order of various operations.

An analogous construction can be carried out at the single-particle level by decomposing the generalized propagator as \cite{Grebenkov20}
\begin{equation}
\calP(\x,z,s|\x_0)=G(\x,s|\x_0)+\calF(\x,z,s|\x_0),
\end{equation}
with
\begin{subequations}
\label{homT}
\begin{align}
  &D\nabla^2 \calF(\x,z,s|\x_0)-s\calF(\x,z,s|\x_0)=0,\ \x \in \Omega,\\
 &\nabla \calF(\x,z,s|\x_0) \cdot \n_2+z\calF(\x,z,s|\x_0) =- \nabla G(\x,s|\x_0) \cdot \n\mbox{ for }  \x\in \partial \Omega_2, \\
 &D\nabla \calF(\x,z,s|\x_0) \cdot \n_1=0 \mbox{ for }  \x\in \partial \Omega_1.
\end{align}
\end{subequations} 
The calculation of $\calF(\x,z,s|\x_0)$ proceeds along similar lines to $\calU(\x,z,s)$ except that now $g(\x,s)=-\partial_{\sigma}G(\x,s|x_0)$ in  equation (\ref{mfH0}) and $J_0=0$. This yields the following spectral decomposition of $\calF(\x,z,s|\x_0)$:
 \begin{equation}
 \label{spec}
\calF(\x,z,s|\x_0)=\frac{1}{D}\sum_{n=0}^{\infty} \frac{{\mathcal V}^*_n(\x_0,s){\mathcal V}_n(\x,s)}{\mu_n(s)+z},
\end{equation}
so that \cite{Grebenkov20}
 \begin{equation}
 \label{spec2}
   \PP(\x,\ell,s|\x_0)=G(\x,s|\x_0)\delta(\ell)+\frac{1}{D}\sum_{n=0}^{\infty}  {\mathcal V}_n^*(\x_0,s){\mathcal V}_n(\x,s) \e^{-\mu_n(s) \ell}.
  \end{equation}

 \subsection{Steady-state solution and the accumulation time}
 
 One major difference between the multiparticle solution (\ref{mspec2}) and the single particle solution (\ref{spec2}) is that the former converges to a non-trivial steady state when $J_0>0$. This makes sense intuitively, since the external flux of particles with zero local time at $\partial \Omega_1$ compensates for the loss of particles with zero local time due to encounters with the surface $\partial \Omega_2$. Mathematically speaking, the result follows by noting that $\lim_{s\rightarrow 0}G(\x',s|\x)\neq 0$, whereas $\lim_{s\rightarrow 0}sG(\x',s|\x)= 0$. Hence, multiplying equation (\ref{solU}) by $s$ and taking the small-$s$ limit shows that
  \begin{equation}
 \label{stU}
  U^*(\x,\ell)=\frac{J_0}{D}\sum_{n=0}^{\infty} \Delta_n^*(0) {\mathcal V}_n(\x,0) \e^{-\mu_n(0) \ell}.
  \end{equation}
 Now suppose that we introduce the stopping local time distribution $\Psi(\ell)$. The multiparticle version of the marginal probability density $p(\x,t|\x_0)$ is 
\begin{equation}
u(\x,t)=\int_0^{\infty}\Psi(\ell)U(\x,\ell,t)d\ell.
\end{equation}
It immediately follows that there exists a steady-state concentration
\begin{equation}
u^*(\x)=\lim_{t\rightarrow \infty}  \int_0^{\infty}\Psi(\ell)U(\x,\ell,t)d\ell=\int_0^{\infty}\Psi(\ell)U^*(\x,\ell)d\ell,
\end{equation}
provided that the integral with respect to $\ell$ is finite. Substituting for $U^*$ using equation (\ref{stU}) and reversing the order of summation and integration, we have
\begin{equation}
u^*(\x)=\frac{J_0}{D}\sum_{n=0}^{\infty} \Delta_n^*(0) {\mathcal V}_n(\x,0)\widetilde{\Psi}(\mu_n(0)).
\end{equation}

Let us introduce an ordering of the eigenvalues of the Dirichlet-to-Neumann operators according to
$\mu_0(s)<\mu_1(s) \leq \mu_2(s) \ldots$.
The principal eigenvalue is typically non-degenerate. Setting $\x\rightarrow \x'$ and $\x_0\rightarrow \x$ in equation (\ref{nabG}), integrating with respect to $\x'\in \Omega$ and using the divergence theorem gives
\begin{equation}
D\int_{\Omega_2}\partial_{\sigma'}G(\x',s|\x)d\x' -s\int_{\Omega} G(\x',s|\x)d\x'=-1.
\end{equation}
It follows that
\begin{equation}
D\partial_{\sigma}\int_{\Omega_2}\partial_{\sigma'}G(\x',s|\x)d\x' -s\partial_{\sigma}\int_{\Omega} G(\x',s|\x)d\x'=0.
\end{equation}
so taking the limit $s\rightarrow 0$, we have
\begin{equation}
-\lim_{s\rightarrow 0} D\partial_{\sigma}\int_{\Omega_2}\partial_{\sigma'}G(\x',s|\x)d\x' =0.
\end{equation}
In other words, there exists a zero eigenvalue $\mu_0(0)=0$ with corresponding eigenvector $v_0(0)=1$.
We thus deduce that a necessary condition for the existence of a steady-state solution is $\widetilde{\Psi}(0)<\infty$. Moreover, using integration by parts, 
\[ \widetilde{\Psi}(0)=\int_0^{\infty} \Psi(\ell)d\ell=[\ell \Psi(\ell)]_0^{\infty}-\int_0^{\infty}\ell \Psi'(\ell)d\ell = \int_0^{\infty}\ell\psi(\ell)d\ell=-\widetilde{\psi}'(0).\]
 Hence, $u^*(\x)$ only exists if $\psi(\ell)$ has a finite first moment.

 Given the solution $u^*(\x)$, we can quantify the rate of relaxation to steady state in terms of an accumulation time. Let
\begin{equation}
\label{accu}
Z(\x,t)=1-\frac{u(\x,t)}{u^*(\x)}
\end{equation}
be the fractional deviation of the concentration from steady state.  (We suppress the explicit dependence on the initial position $\x_0$.) Assuming that there is no overshooting, $1-Z(\x,t)$ can be interpreted as the fraction of the steady-state concentration that has accumulated at $\x$ by time $t$. It follows that $-\partial_t Z(\x, t)dt$ is the fraction accumulated in the interval $[t,t+dt]$. The accumulation time $T_{\rm acc}(\x)$ is then defined as \begin{equation}
\label{accu2}
T_{\rm acc}(\x)=\int_0^{\infty} t\left (-\frac{\partial Z(\x,t)}{\partial t}\right )dt=\int_0^{\infty} Z(\x,t)dt.
\end{equation}
Laplace transforming equation (\ref{accu}) with respect to $t$ and using the identity $u^*(\x)=\lim_{s\rightarrow 0}s\widetilde{u}(\x,s)$,
we have
\[s\widetilde{Z}(\x,s)=1-\frac{s\widetilde{u}(\x,s)}{u^*(\x)}\]
and, hence\footnote{Note that $ T_{\rm acc}(\x)\neq \int_0^{\infty} \Psi(\ell)T_{\rm acc}(\x,\ell)d\ell$, where $T_{\rm acc}(\x,\ell)$ is the accumulation time $ T_{\rm acc}(\x,\ell)$ associated with the relaxation of the generalized propagator.}
\begin{align}
\label{Tuc}
  T_{\rm acc}(\x)=\lim_{s\rightarrow 0} \widetilde{Z}(\x,s) = \lim_{s\rightarrow 0}\frac{1}{s}\left [1-\frac{s\widetilde{u}(\x,s)}{u^*(\x)}\right ] =-\frac{1}{u^*(\x)}
\left .\frac{d}{ds}s\widetilde{u}(\x,s)\right |_{s=0}.
\end{align}
We conclude that the accumulation time $ T_{\rm acc}(\x)$ can be calculated in terms of the Laplace transformed propagator $\U(\x,\ell,s|\x_0)$. In particular, we have
\begin{equation}
s\widetilde{u}(\x,s)=\frac{J_0}{D}\sum_{n=0}^{\infty} \Delta_n^*(s) {\mathcal V}_n(\x,s)\widetilde{\Psi}(\mu_n(s)).
\end{equation}
Differentiating both sides with respect to $s$, reversing the order of summation and differentiation, and taking the limit $s\rightarrow 0$ implies that there will be terms involving both $\widetilde{\Psi}(0)$ and $\widetilde{\Psi}'(0)$. Again using integration by parts, we have
\begin{align*}
 \int_0^{\infty} \ell \Psi(\ell)d\ell=\frac{1}{2}[\ell^2 \Psi(\ell)]_0^{\infty}-\frac{1}{2}\int_0^{\infty}\ell^2 \Psi'(\ell)d\ell = \frac{1}{2}\int_0^{\infty}\ell^2\psi(\ell)d\ell=\frac{1}{2}\widetilde{\psi}''(0).
\end{align*}
Hence, the accumulation time is only well-defined if the first and second moments of $\psi(\ell)$ are finite.

\section{Diffusion in an interval}

Expressing the solution of the general propagator BVP in terms of the spectrum of the associated Dirichlet-to-Neumann operator allowed us to derive necessary conditions for the existence of a steady-state solution and the corresponding accumulation time. In this section we consider an example where the Robin BVP can be solved explicitly without the need for any spectral theory. Consider diffusion in the finite interval $\Omega =[0,L]$ with a partially absorbing boundary at $x=L$. We first calculate the MFPT for absorption at the single particle level by taking a totally reflecting boundary at $x=0$. We then determine the steady-state solution and the accumulation time at the multiparticle level in the case of a constant flux at $x=0$. (Other combinations of boundary conditions could be handled in an analogous fashion.) 

\subsection{Calculation of the MFPT for a single particle}
Performing a double Laplace transform of the 1D version of equations (\ref{Ploc}) yields the propagator BVP 
\begin{subequations}
\label{1D}
\begin{align}
 & D\frac{\partial^2 \calP(x,z,s|x_0)}{\partial x^2}-s\calP(x,z,s|x_0) =- \delta(x-x_0),\ x,x_0\in (0,L),  \\
 & \partial_x\calP(0,z,s|x_0)=0,\quad \partial_x\calP(L,z,s|x_0) =-z \calP(L,z,s|x_0). 
 \end{align}
\end{subequations}
The general solution is
\begin{subequations}
\label{aQ1D}
\begin{align}
\calP(x,z,s|x_0)&= A(z,s) \cosh(\sqrt{s/D} x) + G(x, s| x_0),
\end{align}
\end{subequations}
where $G$ is the 1D Green's function that satisfies equation (\ref{1D}a) with a Neumann boundary condition at $x=0$ and a Dirichlet boundary condition at $x=L$:
\begin{align}
\label{G1D}
  G(x, s| x_0) 
    &= \frac{H(x_0 - x)f(x, s)\widehat{f}(x_0, s) +H(x - x_0)f(x_0, s)\widehat{f}(x, s)}{\sqrt{sD}\cosh(\sqrt{s/D}L)},
\end{align}
where $H(x)$ is the Heaviside function and
\begin{align}
   f(x, s) = \cosh \sqrt{s/D} x,\quad \makebox{and} \quad \widehat{f}(x, s) =\sinh\sqrt{s/D} (L-x). 
\end{align}
Hence, $G$ is the solution to the BVP for a totally absorbing boundary at $x=L$. The unknown coefficient $A(z,s)$ is determined from the Robin boundary condition at $x=L$:\footnote{The solution for $A(z,s)$ could also be obtained by solving an equation of the form (\ref{mfH0}), since the Dirichlet-to-Neumann operator is simply a scalar:
\begin{equation*}
[\L_s f](L,s)=-Df(L,s)\left .\partial_{x}\partial_{x'} G(x',s|x)\right |_{x=x'=L}=f(L,s)\sqrt{\frac{s}{D}} \tanh(\sqrt{s/D}L) .
\end{equation*}
}
\begin{equation}
\label{AA}
A(z,s)=  -\frac{\partial_xG(L,s|x_0)}{\sqrt{s/D}\sinh(\sqrt{s/D}L)+z\cosh(\sqrt{s/D}L)},
\end{equation}
with
\begin{equation}
\label{dG}
\partial_xG(L,s|x_0)=-\frac{1}{D}\frac{\cosh(\sqrt{s/D}x_0)}{\cosh(\sqrt{s/D}L)}.
\end{equation}
It is now straightforward to obtain the  corresponding inverse Laplace transform of $\calP(x,z,s|x_0)$. First, rewrite equation (\ref{AA}) as
\begin{equation}
\label{Ass}
A(z,s)=\frac{A_0(s)}{z+\Lambda(s)},
\end{equation}
with
\begin{align}
\Lambda(s)=\sqrt{s/D } \tanh\sqrt{s/D}L,\quad A_0(s)=-\frac{\partial_xG(L,s|x_0)}{\cosh(\sqrt{s/D}L)}.
\end{align}
It then follows from the general solution (\ref{aQ1D}) that
\begin{equation}
\label{ggg}
\PP(x,\ell,s|x_0)=G(x,s|x_0)\delta(\ell) +A_0(s)\e^{-\Lambda(s)\ell}\cosh(\sqrt{s/D} x) .
\end{equation}

Let $\Psi(\ell)$ be a stopping local time distribution and set
\begin{align}
\label{p1D}
 \p(x,s|x_0)&=\int_0^{\infty}\Psi(\ell) \PP(x,\ell,s|x_0)d\ell = G(x,s|x_0)+A_0(s)\widetilde{\Psi}(\Lambda(s))\cosh(\sqrt{s/D} x).
\end{align}
The corresponding flux through the partially absorbing boundary at $x=L$ is
\begin{align}
\label{J1D}
\widetilde{J}(x_0,s)&=D\int_0^{\infty}\psi(\ell) \PP(L,\ell,s|x_0)d\ell = DA_0(s)\widetilde{\psi}(\Lambda(s))\cosh(\sqrt{s/D} L)\nonumber \\
&=-D\widetilde{\psi}(\Lambda(s))\partial_xG(L,s|x_0) \equiv \widetilde{\psi}(\Lambda(s))\widetilde{J}_{\infty}(x_0,s).
\end{align}
Note that $\widetilde{J}_{\infty}(x_0,s)=-D\partial_xG(L,s|x_0) $ is the flux in the case of a totally absorbing boundary at $x=L$, which corresponds to the case $\widetilde{\psi}(z)=\delta(z)$.
We now use equation (\ref{J1D}) to determine the MFPT for absorption. Differentiating equation (\ref{J1D}) with respect to $s$ and using the 1D version of equation (\ref{MFPT}), we obtain the result
\begin{align}
T(x_0)&=-\left .\frac{\partial}{\partial s}\widetilde{J}(x_0,s)\right |_{s=0}=T_{\infty}(x_0)-\widetilde{\psi}'(0)\Lambda'(0)\widetilde{J}_{\infty}(x_0,0)\nonumber\\
&=T_{\infty}(x_0)-\frac{L}{D}\widetilde{\psi}'(0),
\label{Tfin1D}
\end{align}
where
\begin{equation}
T_{\infty}(x_0)=-\left .\frac{\partial}{\partial s}\widetilde{J}_{\infty}(x_0,s)\right |_{s=0}=\frac{L^2-x_0^2}{2D}
\end{equation}
is the MFPT in the case of a totally absorbing boundary. It immediately follows that if a surface reaction involves a stopping local time distribution with $\widetilde{\psi}'(0)=-\infty$, then the MFPT $T(x_0)$ blows up, indicating that the target is not sufficiently absorbing. In other words, for finite $T(x_0)$ the stopping local time density $\psi(\ell)$ must have a finite first moment.

 \begin{figure}[b!]
  \centering
   \includegraphics[width=8cm]{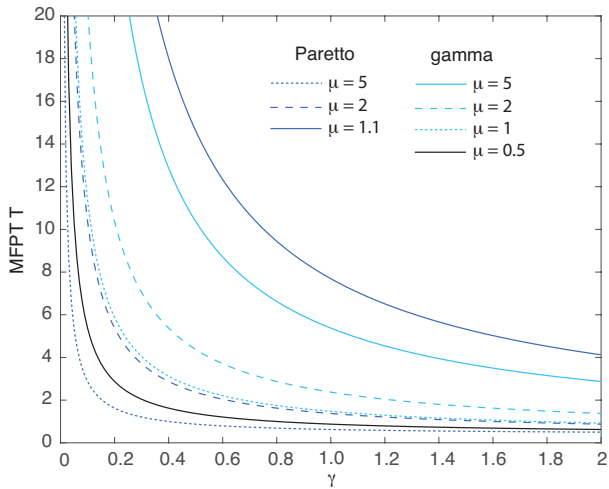}
  \caption{Single-particle diffusion in an interval with a partially absorbing boundary at $x=L$ and a totally reflecting boundary at $x=0$. Plot of MFPT $T(x_0)$ as a function of the distribution parameter $\gamma$ for Paretto-II (light curves) and gamma (dark curves). All curves converge to the MFPT for a totally absorbing boundary, $T_{\infty}(x_0)$, in the limit $\gamma \rightarrow \infty$. Other parameters are $L=D=1$ and $x_0=0.5$.}
  \label{fig2}
  \end{figure}
  
Following previous models of diffusion-mediated absorption \cite{Grebenkov20,Bressloff22a}, we will consider two particular choices for $\psi(\ell)$, namely, the gamma distribution $\psi_{\rm gam}$ and the Pareto-II or Lomax distribution $\psi_{\rm par}$, respectively. The gamma distribution and its associated reactivity function $\kappa(\ell)$ are given by
\begin{equation}
\label{psigam}
\psi_{\rm gam}(\ell)=\frac{\gamma(\gamma \ell)^{\mu-1}\e^{-\gamma \ell}}{\Gamma(\mu)},\quad \kappa(\ell)=\gamma \frac{ (\gamma \ell)^{\mu-1}\e^{-\gamma \ell}}{\Gamma(\mu,\gamma \ell)} ,\ \mu >0,
\end{equation}
where $\Gamma(\mu)$ is the gamma function and $\Gamma(\mu,z)$ is the upper incomplete gamma function:
\begin{equation}
\Gamma(\mu)=\int_0^{\infty}\e^{-t}t^{\mu-1}dt,\quad \Gamma(\mu,z)=\int_z^{\infty}\e^{-t}t^{\mu-1}dt,\ \mu >0.
\end{equation}
Note that $\gamma$ determines the effective absorption rate. In particular, the boundary $x=L$ is non-absorbing in the limit $\gamma\rightarrow 0$ whereas it is totally absorbing in the limit $\gamma \rightarrow \infty$. If $\mu=1$ then $\psi_{\rm gam}$ reduces to the exponential distribution with constant reactivity $\gamma$, that is, $\psi_{\rm gam}(\ell)|_{\mu =1}=\gamma \e^{-\gamma \ell}$. The parameter $\mu$ characterizes the deviation of $\psi_{\rm gam}(\ell)$ from the exponential case. If $\mu <1$ ($\mu>1$) then $\psi_{\rm gam}(\ell)$ decreases more rapidly (slowly) as a function of the local time $\ell$. The Pareto-II (Lomax) distribution and its reactivity function take the form
\begin{equation}
\psi_{\rm par}(\ell)=\frac{\gamma \mu}{(1+\gamma \ell)^{1+\mu}},\quad \kappa(\ell)= \frac{\gamma\mu }{1+\gamma \ell} ,\quad \mu >0,
\end{equation}
Note that $\psi_{\rm gam}(\ell)$ has finite moments for all $\mu >0$, whereas $\psi_{\rm par}(\ell)$ only has finite moments when $\mu >1$. The blow up of the moments when $\mu <1$ reflects the fact that the Pareto-II distribution then has a long tail. The corresponding Laplace transforms are
\begin{equation}
\widetilde{\psi}_{\rm gam}(z)=\left (\frac{\gamma}{\gamma+z}\right )^{\mu},\quad \widetilde{\psi}_{\rm gam}'(z)=-\mu \left (\frac{\gamma}{\gamma+z}\right )^{\mu}\frac{1}{\gamma+z}.
\end{equation}
and
\begin{subequations}
\begin{align}
 \widetilde{\psi}_{\rm par}(z)&=\mu\left (\frac{z}{\gamma}\right )^{\mu}\e^{z/\gamma}\Gamma(-\mu,z/\gamma),\\
  \widetilde{\psi}_{\rm par}'(z)&=\mu\left (\frac{z}{\gamma}\right )^{\mu}\e^{z/\gamma}\left (\left [\frac{\mu}{z}+\frac{1}{\gamma}\right ]\Gamma(-\mu,z/\gamma)+\partial_{z} \Gamma(-\mu,z/\gamma)\right ).
\end{align}
\end{subequations}
Example plots of the MFPT $T(x_0)$ for these two distributions are shown in Fig. \ref{fig2}. Note that in the case of the Paretto-II distribution, the MFPT is only defined when $\mu >1$. It can be seen that all the curves converge asymptotically to the limiting value $T_{\infty}(x_0)$ as $\gamma \rightarrow \infty$. Moreover, in the case of the gamma distribution, the MFPT is an increasing function of $\mu$, whereas the converse holds for Paretto-II.

\subsection{Calculation of the accumulation time for multiple particles}

Consider the following 1D version of the multiparticle BVP (\ref{mPlocLT}):
\begin{subequations}
\label{m1D}
\begin{align}
 & D\frac{\partial^2 \calU(x,z,s)}{\partial x^2}-s\calU(x,z,s) =0,\ x\in (0,L),  \\
 & D \partial_x\calU(0,z,s)=-\frac{J_0}{s},\quad \partial_x\calU(L,z,s) =-z \calU(L,z,s). 
 \end{align}
\end{subequations}
The general solution is
\begin{align}
\label{mQ1D}
\calU(x,z,s)&= A(z,s) \cosh(\sqrt{s/D} x) -\frac{1}{\sqrt{sD}}\frac{J_0}{s}\sinh(\sqrt{s/D} x) .
\end{align}
The unknown coefficient $A(z,s)$ is again determined by the Robin boundary condition at $x=L$:
\begin{equation}
\label{mAA}
A(z,s)=  \frac{\displaystyle J_0}{\displaystyle s}\frac{  z\sinh(\sqrt{s/D}L) /\sqrt{Ds}+D^{-1}\cosh(\sqrt{s/D}L) }{\sqrt{s/D}\sinh(\sqrt{s/D}L)+z\cosh(\sqrt{s/D}L)},
\end{equation}
where $G$ is the 1D Green's function (\ref{G1D}). 
Multiplying the solution (\ref{mQ1D}) by $s$ and taking the limit $s\rightarrow 0$ yields the non-trivial steady-state concentration
\begin{equation}
\calU^*(x,z)=\frac{J_0[L-x]}{D}+\frac{J_0}{zD}.
\end{equation}
Inverting with respect to $z$ then implies
\begin{equation}
U^*(x,\ell)=\frac{J_0[L-x]}{D}\delta(\ell)+\frac{J_0}{D}.
\end{equation}
It follows that the steady-state concentration $u^*(x)$ for a given stopping time distribution $\Psi(\ell)$ is
\begin{equation}
u^*(x)=\int_0^{\infty}\Psi(\ell)U^*(x,\ell)d\ell = \frac{J_0[L-x]}{D}+\frac{J_0}{D}\widetilde{\Psi}(0).
\end{equation}
Such a solution will only exist if $\widetilde{\Psi}(0)<\infty$. Using integration by parts, we see that
\[\widetilde{\Psi}(0)=\int_0^{\infty} \Psi(\ell)d\ell=[\ell \Psi(\ell)]_0^{\infty}-\int_0^{\infty}\ell \Psi'(\ell)d\ell = \int_0^{\infty}\ell\psi(\ell)d\ell.\]
 Hence, $u^*(x)$ exists if and only if $\psi(\ell)$ has a finite first moment. This is consistent with our previous result for the MFPT, namely, if $\psi(\ell)$ has a large tail then absorption is too weak to counterbalance the influx at $x=0$.

 When $u^*(x)$ exists we can quantify the rate of relaxation in terms of the accumulation time. 
The 1D version of equation (\ref{Tuc}) is
\begin{align}
\label{Tuc1Ds}
 T_{\rm acc}(x) =-\frac{1}{u^*(x)}
\left .\frac{d}{ds}s\widetilde{u}(x,s)\right |_{s=0}.
\end{align}
In order to determine $T_{\rm acc}(x)$ we first have to calculate the derivative of $s\calU(x,z,s)$ and then take the limit $s\rightarrow 0$. The simplest way to proceed is to Taylor expand the solution (\ref{mQ1D}) with respect to $s$. In particular,
\begin{align*}
 &sA(z,s)=J_0  \left \{\frac{z}{\sqrt{Ds}}\left (\sqrt{s/D}L +(\sqrt{s/D}L)^3/6  \right )+\frac{1}{D} \left (1 +(\sqrt{s/D}L)^2/2 \right ) \right \}+\ldots\\
 &\qquad \times\left  \{\sqrt{s/D}\left (\sqrt{s/D}L +(\sqrt{s/D}L)^3/6 +\ldots\right )+z\left (1 +(\sqrt{s/D}L)^2/2 +\ldots\right )\right \}^{-1}\\
 &=J_0\left \{ (1+zL)/D+sL^2(zL/3+1)/2D^2  \right \} \left \{ z+sL(1+zL/2)/D\right \}^{-1}+\ldots\\
 &=\frac{s}{zD}+\frac{J_0}{zD}(1+zL)-s\frac{J_0}{D^2}\left \{\frac{L^3}{3}+\frac{L^2}{z}+\frac{L}{z^2}\right \}+O(s^2).
\end{align*}
Hence
\begin{align}
 -\left .\frac{d}{ds}sA(z,s)\cosh(\sqrt{s/D}x)\right |_{s=0}=\frac{J_0}{D^2}\left \{\frac{L^3}{3}-\frac{x^2L}{2}+\frac{2L^2-x^2}{2z}+\frac{L}{z^2}\right \}.
\end{align}
Moreover,
\begin{align}
\left .\frac{d}{ds}\frac{J_0}{\sqrt{sD}}\sinh(\sqrt{s/D}x)\right |_{s=0}=\frac{J_0}{D^2}\frac{x^3}{6}.
\end{align}
Combining the last two equations thus gives
\begin{align}
 -\left .\frac{d}{ds}s\calU(x,z,s)\right |_{s=0}=\frac{J_0}{D^2}\left \{\frac{L^3}{3}+\frac{x^3}{6}-\frac{x^2L}{2}+\frac{2L^2-x^2}{2z}+\frac{L}{z^2}\right \}.
\end{align}
Inverting with respect to $z$ implies that
\begin{align}
 -\left .\frac{d}{ds}s\U(x,\ell,s)\right |_{s=0}=\frac{J_0}{D^2}\left \{\left [\frac{L^3}{3}+\frac{x^3}{6}-\frac{x^2L}{2}\right ]\delta(\ell)+\frac{2L^2-x^2}{2}+\ell L\right \}.
\end{align}

\begin{figure}[t!]
\centering
  \includegraphics[width=11cm]{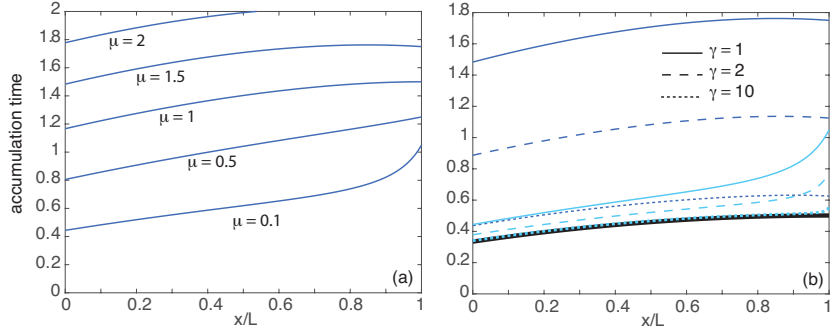}
  \caption{Accumulation time $T_{\rm acc}(x)$ for diffusion in a finite interval $[0,L]$, with a constant flux at $x=0$ and a partially absorbing boundary at $x=L$. The stopping local time density is taken to be the gamma distribution (\ref{psigam}) with parameters $\mu,\gamma$. (a) Plot of the accumulation time as a function of position $x$ for $\gamma=1$ and various $\mu$. (b) Corresponding plots for various $\gamma$ with $\mu=0.1$ (light curves) and $\mu=1.5$ (dark curves). The thick line corresponds to a totally absorbing boundary at $x=L$ ($\gamma \rightarrow \infty$). We have also set $D=1$ and $L=1$.}
  \label{fig3}
\end{figure}

Finally, multiplying both sides by $\Psi(\ell)/u^*(x)$ and integrating with respect to $\ell$ leads to the following explicit expression for the accumulation time:
\begin{align}
\label{tact}
 T_{\rm acc}(x)&=\frac{1}{D[L-x+\widetilde{\Psi}(0)]}\\
 &\quad \times \left \{\frac{L^3}{3}+\frac{x^3}{6}-\frac{x^2L}{2}+\frac{2L^2-x^2}{2}\int_0^{\infty}\Psi(\ell)d\ell+L\int_0^{\infty}\ell \Psi(\ell)d\ell\right \}.\nonumber
\end{align}
Note that in the case of a totally absorbing boundary $\partial \Omega_1$ this reduces to
\begin{align}
T^{\infty}_{\rm acc}(x)&=\frac{1}{D[L-x]}\left \{\frac{L^3}{3}+\frac{x^3}{6}-\frac{x^2L}{2}\right \}=\frac{2(L+x)L-x^2}{6D}.
\end{align}
Again using integration by parts, we have $\int_0^{\infty} \Psi(\ell)d\ell=-\widetilde{\psi}'(0)$ and 
\begin{align*}
 \int_0^{\infty} \ell \Psi(\ell)d\ell=\frac{1}{2}[\ell^2 \Psi(\ell)]_0^{\infty}-\frac{1}{2}\int_0^{\infty}\ell^2 \Psi'(\ell)d\ell = \frac{1}{2}\int_0^{\infty}\ell^2\psi(\ell)d\ell=\frac{1}{2}\widetilde{\psi}''(0).
\end{align*}
Equation (\ref{tact}) then implies that the accumulation time is only well-defined if the first and second moments of $\psi(\ell)$ are finite, consistent with the general result obtained in section 3. For the sake of illustration, let $\psi(\ell)$ be the gamma distribution (\ref{psigam}) so that
\[\widetilde{\psi}'(0)=-\frac{\mu}{\gamma},\quad \widetilde{\psi}''(0)=\frac{\mu(\mu+1)}{\gamma^2}.\]
In Fig. \ref{fig3} we show example plots of $T_{\rm acc}(x)$ as a function of $x$ for various values of the parameters $\mu,\gamma$. In Fig. \ref{fig3}(a) it can be seen that the accumulation time is increased (decreased) for $\mu>1$ ($\mu<1$) when compared to the exponential case ($\mu=1$). Moreover, $T_{\rm acc}(x)$ is a non-monotonic function of $x$ when $\mu >1$. Fig. \ref{fig3}(b) shows that the accumulation time converges to the case $T_{\rm acc}^{\infty}$ in the limit $\gamma \rightarrow \infty$.

\section{Diffusion in a $d$-dimensional spherical shell}

As our second example, we consider the multiparticle steady-state solution for a spherical shell $\Omega =\{\x\in \R^d\,|\, R_1 <  |\x| <R_2\}$ with $\partial \Omega_j= \{\x\in \R^d\,|\,  |\x| =R_j\}$, $j=1,2$. 
Introducing spherical polar coordinates in the BVP (\ref{mPlocLT}) gives
  \begin{subequations}
  \label{sph}
\begin{align}
    &D\frac{\partial^2\calU}{\partial \rho^2} + D\frac{d - 1}{\rho}\frac{\partial \calU}{\partial \rho} -s\calU(\rho,z, s) = 0, \ R_1<\rho <R_2,\\ 
   & \left . \frac{\partial }{\partial \rho}\calU(\rho,z,s)\right |_{\rho=R_1}=-\frac{J_0}{s},\quad \left . \frac{\partial }{\partial \rho}\calU(\rho,z,s)\right |_{\rho=R_2}=-z\calU(R_2,z,s).
  \end{align}
\end{subequations}
 Equations of the form (\ref{sph}) can be solved in terms of modified Bessel functions \cite{Redner01}. The general solution is
  \begin{align}
\label{qir}
     \calU(\rho,z, s) = B(z,s) \rho^\nu I_\nu(\alpha \rho)  + C(z,s)\rho^\nu K_\nu(\alpha \rho) , \ \rho \in (R_1, R_2),
\end{align}
with $\nu = 1 - d/2$ and $\alpha=\sqrt{s/D}$. In addition, $I_{\nu}$ and $K_{\nu}$ are modified Bessel functions of the first and second kind, respectively. 
The unknown coefficients $B(z,s)$ and $C(z,s)$ are determined from the boundary conditions (\ref{sph}b). Setting
\begin{equation}
F_I(\rho,s)=\rho^\nu I_\nu(\alpha \rho),\quad F_K(\rho,s)=\rho^\nu K_\nu(\alpha \rho),
\end{equation}
\begin{subequations}
we have
\begin{equation}
\label{CB1}
 B(z,s) F_I'(R_2,s)+C(z,s)F_K'(R_2,s)=-z[B(z,s) F_I(R_2,s)+C(z,s)F_K(R_2,s)],
\end{equation}
and
\begin{align}
 B(z,s)F_I'(R_1,s) +C(z,s)F_K'(R_1,s)&=-\frac{J_0}{s}
\label{CB2}
\end{align}
\end{subequations}
Here $'$ denotes differentiation with respect to $\rho$. 
Equation (\ref{CB1}) shows that
\begin{equation}
B(z,s)=-\frac{F_K'(R_2,s)+zF_K(R_2,s)}{F_I'(R_2,s)+zF_I(R_2,s)}C(z,s)\equiv -\Lambda(z,s)C(z,s).
\end{equation}
Substituting into equation (\ref{CB2}) and rearranging yields 
\begin{equation}
\label{C}
C(z,s)= -\left (F_K'(R_1,s)-F_I'(R_1,s)\Lambda(z,s)\right )^{-1}\frac{J_0}{s}
\end{equation}
Combining our various results yields the following solution for the Laplace transformed generalized concentration:
\begin{align}
 & \calU(\rho,z,s)= -\frac{F_K(\rho,s)-F_I(\rho,s)\Lambda(z,s)}{F_K'(R_1,s)-F_I'(R_1,s)\Lambda(z,s)}\frac{J_0}{s}\nonumber \\
  &=- \frac{F_K(\rho,s)[F_I'(R_2,s)+zF_I(R_2,s)]-F_I(\rho,s)[F_K'(R_2,s)+zF_K(R_2,s)]}{F_K'(R_1,s)[F_I'(R_2,s)+zF_I(R_2,s)]-F_I'(R_1,s)F_K'(R_2,s)+zF_K(R_2,s)]}\frac{J_0}{s}\nonumber \\
  &=-\frac{\Theta_1(\rho,s)+z\Theta_2(\rho,s)}{\Theta_1'(R_1,s)+z\Theta_2'(R_1,s)}\frac{J_0}{s},
 \end{align}
 where
 \begin{subequations}
 \begin{align}
 \Theta_1(\rho,s)&=F_K(\rho,s)F_I'(R_2,s)-F_I(\rho,s)F_K'(R_2,s),\\  \Theta_2(\rho,s)&=F_K(\rho,s)F_I(R_2,s)-F_I(\rho,s)F_K(R_2,s).
 \end{align}
 \end{subequations}
 
 \begin{figure}[t!]
\centering
  \includegraphics[width=11cm]{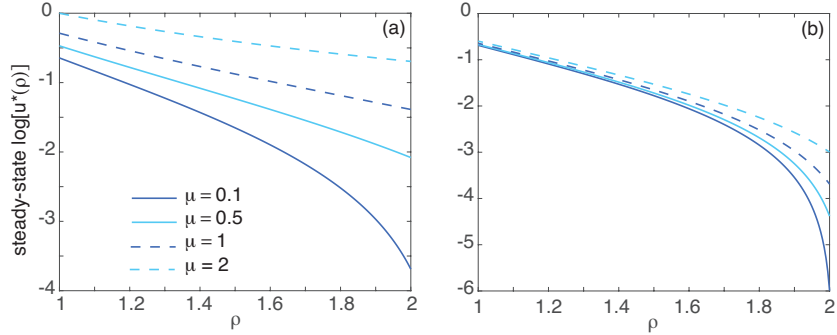}
  \caption{Log-linear plot of steady-state concentration $u^*(\rho)$ as a function of the radius $\rho$ for a 3D spherical shell with inner radius $R_1=1$ and outer radius $R_2=2$. We assume that the stopping time density is a gamma distribution with parameters $\gamma,\mu$. (a) $\gamma=1$; (b) $\gamma=10$. We have also set $D=1$ and $J_0=1$.}
  \label{fig4}
\end{figure}

 In order to obtain the inverse Laplace transform with respect to $z$, we rewrite the propagator as
 \begin{align}
   \calU(\rho,z,s)&=-\left [\frac{\Theta_1(\rho,s)}{\Theta_2'(R_1,s)}\frac{1}{z+\Theta(s)}+\frac{\Theta_2(\rho,s)}{\Theta_2'(R_1,s)} \left \{ 1-\frac{\Theta(s)}{z+\Theta(s)}\right \}\right ]\frac{J_0}{s},
 \end{align}
 where
 \begin{equation}
 \Theta(s)=\frac{\Theta_1'(R_1,s)}{\Theta_2'(R_1,s)}.
 \end{equation}
The inverse Laplace transform is then simply
\begin{align}
   \U(\rho,\ell,s)&=-\left [\frac{\Theta_2(\rho,s)}{\Theta_2'(R_1,s)} \delta(\ell)+\left (\frac{\Theta_1(\rho,s)}{\Theta_2'(R_1,s)} -\Theta(s) \frac{\Theta_2(\rho,s)}{\Theta_2'(R_1,s)}\right )\e^{-\Theta(s)\ell}\right ]\frac{J_0}{s},
 \end{align}
Given a stopping local time distribution $\Psi(\ell)$, the corresponding marginal concentration is 
\begin{align}
  \widetilde{u}(\rho,s)=-\left [\frac{\Theta_2(\rho,s)}{\Theta_2'(R_1,s)} +\left (\frac{\Theta_1(\rho,s)}{\Theta_2'(R_1,s)} -\Theta(s) \frac{\Theta_2(\rho,s)}{\Theta_2'(R_1,s)}\right )\widetilde{\Psi}(\Theta(s))\right ]\frac{J_0}{s}.
 \label{ppp}
\end{align}
Finally, multiplying both sides by $s$ and taking the limit $s\rightarrow 0$ with $\Theta(s) \rightarrow 0$ yields the steady-state concentration
\begin{align}
u^*(\rho)=-\left [\lim_{s\rightarrow 0}\frac{\Theta_2(\rho,s)}{\Theta_2'(R_1,s)} + \lim_{s\rightarrow 0}\frac{\Theta_1(\rho,s)}{\Theta_2'(R_1,s)} \widetilde{\Psi}(0)\right ]J_0
\end{align}
Again we require that the density $\psi(\ell)$ has a finite first moment. As expected, the steady state is a monotonically decreasing function of $\rho$. In Fig. \ref{fig4} we plot $\log u^*(\rho)$ as a function of $\rho$ for the gamma distribution, where $\widetilde{\Psi}(0)=\mu/\gamma$. We also take $d=3$. It can be seen that for $\gamma=1$ the steady-state decays less steeply as $\mu$ increases. On the other hand, the dependence on $\mu$ vanishes in the limit $\gamma \rightarrow \infty$, since the second term in the square brackets becomes zero. Similar results hold for a 2D shell.
Note that the solution (\ref{ppp}) could also be used to calculate the accumulation time, although the algebra is considerably more involved than the 1D case. Nevertheless, the qualitative behavior is similar.

\section{Discussion}

In this paper we considered the relationship between single-particle (microscopic) and multiparticle (macroscopic) interpretations of diffusion within the context of partially absorbing boundaries. Using an encounter-based model of single-particle diffusion, we constructed a BVP for the concentration of a population of particles in an extended phase space consisting of both particle position and boundary local time. Absorption was then incorporated by introducing a random stopping condition for the local time. In addition, the loss of particles through surface absorption was counterbalanced by external fluxes, resulting in a nontrivial steady-state. Solving the BVP in Laplace space allowed us to derive general expressions for the steady-state concentration and the associated accumulation time, which were based on the spectral decomposition of an associated Dirichlet-to-Neumann operator. This was then used to derive necessary conditions for the existence of a steady-state solution and a finite accumulation time. We illustrated the theory by considering diffusion in a finite interval and in a $d$-dimensional spherical shell.
At the single-particle level, we have recently analyzed a narrow capture problem for diffusion in a singularly perturbed domain containing several small spherical targets or traps \cite{Bressloff22b}. The boundary of each trap was taken to be partially absorbing. Using a mixture of matched asymptotic analysis and Green's function methods, we solved the resulting BVP and calculated the splitting probabilities and conditional MFPTs. It would be interesting to develop a multiparticle version of the narrow escape problem, in order to investigate the existence of steady-state solutions and the associated accumulation times.

\begin{footnotesize}

\end{footnotesize}

\end{document}